\documentclass[two column]{aa}
\usepackage{graphicx}
\usepackage{txfonts}
\begin{document}
   \title{A snapshot of the inner dusty regions of a R~CrB-type
variable\thanks{Based on observations collected
           with the VLTI/MIDI instrument at Paranal Observatory, ESO (Chile) --
           Programme 75.D-0660.}}
   
   \author{I.C. Le\~ao\inst{1,2} \and P. de Laverny\inst{1}
           \and O. Chesneau\inst{3} 
           \and D. M\'ekarnia\inst{1}\thanks{Winterover 2007 at Concordia base (IPEV-PNRA), Dome C, Antarctic}
           \and J.R. De Medeiros\inst{2}}

   \offprints{P. de Laverny (laverny@obs-nice.fr)}
   
   \institute{Observatoire de la C\^ote d'Azur,
              Dpt Cassiop\'ee, CNRS - UMR 6202, BP 4229, 06304 Nice Cedex 4, France
         \and Departamento de F\'isica, Universidade Federal do Rio Grande do Norte,
              59072-970 Natal, RN, Brazil
         \and Observatoire de la C\^ote d'Azur, Dpt Gemini - CNRS - UMR 6203,
              Avenue Copernic, F-06130, Grasse, France         
              }
   \date{Received Month Day, Year; accepted Month Day, Year}

\abstract
{R~Coronae Borealis (R~CrB) variable stars are suspected to 
sporadically eject optically thick dust clouds causing, when one of them lies
on the line-of-sight, a huge brightness decline in visible light.
Direct detections with 8-m class adaptive optics of such clouds located at about 0.2-0.3~arcsec from the centre ($\sim$1\,000 stellar radii) were recently reported by de Laverny \& M\'ekarnia (2004) for RY~Sgr, the brightest R~CrB of the southern hemisphere. 
}
{Mid-infrared interferometric observations of RY~Sgr allowed us to explore the circumstellar regions much closer to the
central star ($\sim$20-40~mas) in order to look for the
signature of any heterogeneities and to characterize them.}
{Using the VLTI/MIDI instrument, five dispersed visibility curves in 
the N band were recorded 
in May and June 2005 with different projected baselines oriented
towards two roughly perpendicular directions.
The large spatial frequencies visibility curves exhibit a sinusoidal shape
whereas, at shorter spatial frequencies visibility curves follow a Gaussian decrease. 
These observations are well interpreted with a geometrical
model consisting in a central star surrounded by an extended circumstellar
envelope in which one bright cloud is embedded.}
{Within this simple geometrical scheme, the inner 110\,AU 
dusty environment of RY\,Sgr is dominated at the time of observations by a single 
dusty cloud which, at 10$\mu$m
represents $\sim$10\% of the total flux of the whole system, slightly less that the star flux. The cloud is located at about 100~stellar 
radii (or $\sim$30 AU) from the centre toward the East-North-East direction 
(or the symmetric direction
with respect to centre) within a circumstellar
envelope which FWHM is about 120~stellar radii.
This first detection of a cloud so close 
to the central star, supports the classical scenario of the
R~CrB brightness variations in the optical spectral domain and demonstrates the feasibility of a temporal monitoring of the dusty environment of this star at a monthly scale.
}
{}


   \maketitle
%

\section{Introduction}

R~Coronae Borealis (R~CrB) variable stars are 
Hydrogen-deficient supergiants exhibiting erratic variabilities.
Their visual light-curve is indeed characterized by unpredicted
declines of up to 8 magnitudes with time-scale of weeks,
the return to normal light being much slower (see Clayton, \cite{cla96}, for
a review).
It has been accepted for decades that such fading could be due
to obscurations of the stellar surface by newly formed dusty clouds.
Over the years, several indices
confirming this scenario were reported although
no direct detections of such clouds have been performed.
Recently, NACO/VLT near-infrared adaptive optics observations by
de Laverny \& M\'ekarnia (\cite{lav04}, Paper~1 hereafter) detected
clear evidences of the presence of such clouds around RY~Sgr,
the brightest R~CrB variable in the southern hemisphere.
New informations about the inner circumstellar regions of these stars
were derived as, for instance, (i) several bright and 
large dusty clouds are present 
around R~CrB variables, (ii) they have been detected in any directions 
at several hundred stellar radii of RY~Sgr, 
(iii) they can be as bright as 2\% of the stellar flux in the
near-infrared,...
This was the first direct confirmation of the standard scenario
explaining R~CrB variable stars light variations by
the presence of heterogeneities in their inner
circumstellar envelope.

However, the precise location of the formation of such
dust clouds is still unclear. The brightest cloud detected in Paper~1
indeed lies at several hundred stellar radii from the center
but it was certainly formed much closer. This cannot help to disentangle 
between the two commonly proposed scenarios 
regarding the location of the dust formation in the vicinity of R~CrB variable
stars in order to explain their fadings:
either the dust is formed very close to the
stellar surface ($\sim 2~$R$_*$ or even less) above large
convection cells or it is formed
in more distant regions at $\sim 20~$R$_*$ where the temperature
is lower to form dust more easily 
(see Clayton \cite{cla96} and Feast \cite{feast}). 
Nothing is also known about the physical and chemical properties of these
clouds, witnesses of nucleation processes 
in a rather hot environment. Indeed, the 
temperature of the layers where they are formed is certainly too high for 
classical dust formation theories and departures from the chemical and 
thermodynamical equilibria have therefore to be invoked.

In the present Letter, we report on the interferometric detection
of a dusty cloud in the very inner environment of RY~Sgr,
i.e. in regions located about one-tenth of the
distance reported in Paper~1.
We present in Sect.~2 the observations and their reduction.
The interpretation of the collected visibility curves  
with a geometrical model is
described in Sect.~3. We then validate the adopted model 
with respect to more complex geometries of the circumstellar environment
of RY~Sgr. In the last section, we finally discuss our results within 
the framework of our understanding of R~CrB variable stars variability.


\section{Observations and data reduction}

N-band interferometric data of \object{RY~Sgr} were collected in 2005
with the VLTI/MID-infrared Interferometric instrument
(MIDI; Leinert et al.~\cite{lei03}).
Seven runs were executed, using two different telescope pairs (UT1-UT4 and 
UT3-UT4) and five different baselines.
Their orientations are shown in Fig.~\ref{figvismono} (top panel).
All observing runs were collected under rather good atmospheric conditions.
The observations were executed in the so-called High-Sens mode, with 4 templates: acquisition, fringe search,
fringe tracking, and photometry.
These templates provide, in addition to
the dispersed (7.5--13.5~$\mu$m) correlated flux visibilities,
N-band adaptive-optics corrected acquisition images 
and spectro-photometric data for each baseline.
We used the grism for wavelength dispersion ($\lambda / \Delta\lambda$=230) and \object{HD~177716} as interferometric,
spectrophotometric and imaging calibrator. The observing log is summarized in Table~\ref{tabobs}.
\begin{table}
 \caption{MIDI observations log of RY~Sgr and its calibrator.}
 \centering
 \label{tabobs}
 \begin{tabular}{ l c | c r l | c }
  \hline \hline
           &       & \multicolumn{3}{c |}{RY~Sgr}                       & HD 177716   \\
  \hline
  Base  & Date  & UT Time     & \multicolumn{2}{c |}{Proj. baseline} & UT Time     \\
      &       &             & length        & PA                   &             \\
           &       &             & (m)           & (deg)                &             \\
  \hline
  UT3-4    &  2005 May 26 & $e$~~~06:01-06:11 & 57      & 98                   & 04:54-05:04 \\
           &       & $g$~~~10:11-10:21 & 57      & 135                  & 10:36-10:46 \\
  UT1-4    & 2005 June 25 & $a$~~~03:01-03:10 & 122     & 34                   & 02:28-02:36 \\
           &       & $b$~~~03:12-03:20 & 123       & 36                   & 03:44-03:52 \\
           &       & $c$~~~06:18-06:26 & 128       & 65                   & 05:52-06:00 \\
           & 2005 June 26 & $d$~~~06:42-06:50 & 125       & 68                   & 06:19-06:28 \\
  UT3-4    & 2005 June 28 & $f$~~~05:28-05:36 & 62      & 110                  & 05:05-05:13 \\
  \hline
 \end{tabular}
 Note: For baseline labels {\it a-g}, see Fig.~\ref{figvismono}.
\end{table}

The observations have been reduced using the MIA software\footnote{http://www.mpia-hd.mpg.de/MIDISOFT/}.
The fringe tracking of baselines $a$ and $c$ was not satisfactory and good scans were carefully selected based on the histogram of Fourier-Amplitude. 
The error on the visibilities were estimated by
examining level and shape fluctuations of several calibrator visibility 
curves collected $\pm 2$~hours around every RY~Sgr observations.
We note that the uncertainties on the visibilities are mostly achromatic and 
are dominated by the fluctuations of the photometry between 
the fringe and photometric
measurements. The error on the spectral shape of the visibility is smaller 
than 2\% of the visibilities and is considered in the following as an 
important constraint of the model fitting process.
The visibility curves are shown in Fig.~\ref{figvismono}. The MIDI spectrum of RY~Sgr was calibrated 
using a template of HD~177716 (Cohen et al.~\cite{coh99}) and a mean flux error of 12\% was estimated from the level fluctuations of all collected spectra.
The MIDI spectrum of RY~Sgr is similar to the ISO one, but about 25\% fainter
(probably due to photometric variations and the smaller field-of-view
of MIDI).
Both spectra exhibit a slow decline between 7.5 and 13.5$\mu$m,
compatible with a continuum dominated by hot dust emission.
Finally, we processed the 8.7~$\mu$m acquisition images of a single 8m
telescope (the FWHM of the beam is 225~mas) by using a shift and add procedure, and found that RY~Sgr is unresolved.
Moreover, no structures were resolved in these N-band images with a 
field-of-view of $\sim$2" and a rather low dynamics ($\sim$20-40).


\section{Interpretation of the visibility curves \label{secres}}

\begin{figure}
   \centering
   \includegraphics[width=9.3cm,height=13.5cm]{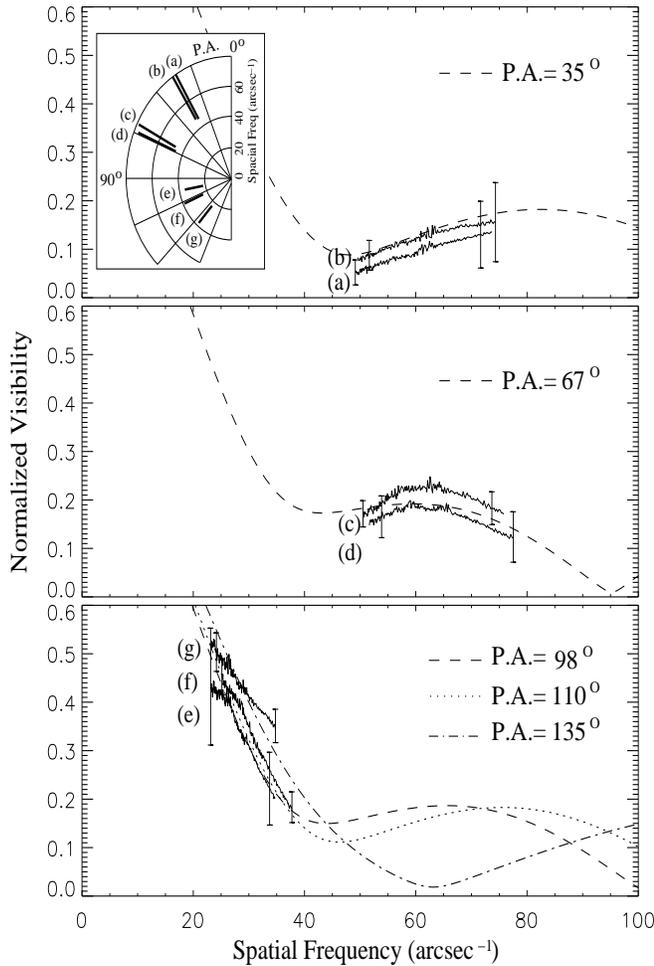}
   \caption{Observed visibilities as a function of spatial frequency and PA.
            The chart of the observed baselines, representing their projected lengths and PAs,
            is shown at the top-left corner, with the same labels as in Table~\ref{tabobs}.
            The different PAs are also indicated in each panel.
              The total error bars on the visibilities are shown at each 
              curve extremities in order to illustrate their 
              mean variations as a function of wavelength.
         The non-solid curves represent PA sections of the theoretical
            fit with parameters described in Sect.~\ref{secres} and neglecting
            any morphological variations with wavelength.}
      \label{figvismono}
    \end{figure}

Figure~\ref{figvismono} shows the visibility curves as a function of spatial frequency
and position angle (PA).
Let us recall that the shapes of these curves  are determined both by an apparently monotonic change of the object geometrical characteristics between 7.5 and 13.5~$\mu$m and the linear decrease of the resolving power of the interferometer with wavelength.
Nevertheless, in a first-order analysis, we neglect any 
variations with wavelength of the source geometry and tried to fit the curves with simple monochromatic geometric models.
This approach gives us fundamental constraints to determine the global morphology of this object in the N-band.
Morphological variations with wavelength will be discussed hereafter.

We identify in the observed visibility curves two main signatures.
At low spatial frequencies and PA~$>$~90\degr \ (baselines labeled 
$e$ to $g$), the visibility curves 
have a Gaussian shape while, at higher spatial frequencies 
and PA~$<$~90\degr \ (projected baselines labeled $a$ to $d$),
they follow a sinusoidal shape, typical of a two-component signature.
These interferometric signatures can be easily interpreted 
with a geometrical model consisting in a
central star and a cloud (the sinusoidal component in the Fourier space),
embedded within an extended circumstellar envelope (the Gaussian component).

We calculated theoretical visibility curves for this geometrical model
and adjusted its parameters to obtain the most reasonable fit.
Let us recall that the errors on the shape of the visibility curves 
are rather small compared to those on their level. We have therefore
given a larger weight for the curve shapes than for their levels 
to adjust the fit. In this way, we estimated a separation for the cloud
of 16$\pm 1$~mas from the central star
and a PA of 75\degr$\pm$10\degr (modulo 180\degr,
because of the central symmetry of the $u$-$v$ plane).
We estimated the FWHM of the Gaussian CSE to be 18$\pm 3$~mas.
The central star contributes
to 10\%$\pm 2\%$ of the total N-band flux
of the whole system, close to the cloud
contribution (8\%$\pm 2\%$).
The best fit of the visibility curves (with no morphological
variations with wavelength, in particular the relative fluxes between its
three components) for these parameter values
are shown in Fig.~\ref{figvismono}.

\begin{figure}
   \centering
   \includegraphics[width=8.6cm]{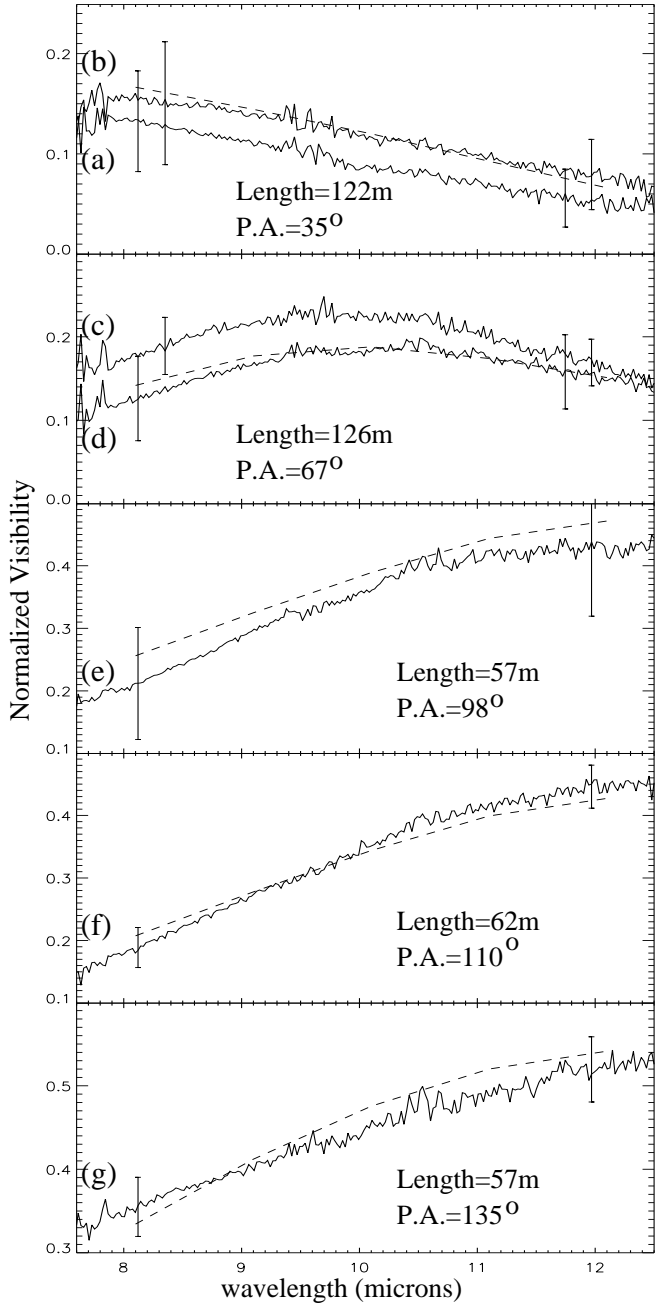}
     \caption{Visibility curves as a function of wavelength for the different
              baselines.
              Their respective lengths and PAs are indicated in each panel.
              The total error bars on the visibilities are shown at each 
              curve extremities in order to illustrate their 
              mean variations as a function of wavelength.
              Best theoretical fits (taking into account any morphological
              variations with wavelength, see end of Sect.~3) 
              are shown as dashed lines.}
      \label{figviswl}
    \end{figure}

In a more detailed analysis, we considered possible spectral variations
of the model parameters.
The modeled visibilities are displayed as a function of wavelength
for each observed baseline in Fig.~\ref{figviswl}.
We started by assuming that the estimated distance 
between the cloud 
and the centre as its PA 
have to be constant within the full N-band. The parameters that are allowed to 
vary in the observed wavelength range were the
FWHM of the envelope and the relative fluxes of the star, the cloud, and the
envelope, using the first order analysis presented above as a starting point.
Very good fits were found with parameter values close to those given in the global analysis,
confirming this first solution. 
In addition, we also found that the CSE FWHM slightly grows from 17 to 19~mas
(within uncertainty of $\pm$3~mas) toward larger 
wavelengths, whereas no significant variation of the stellar and cloud fluxes
were observed. Actually, although smaller errors on the visibility curves would help,
there is a degeneracy when estimating any spectral
variation of the flux of each component with the simple geometrical 
model considered above.
For instance, we see in Fig.~\ref{figvismono} that
the shapes of the visibility curves $c$ and $d$
are slightly more bent than their theoretical fit.
They could be better fitted
either by slightly decreasing the stellar flux
 with increasing wavelength or 
by slightly increasing the cloud  or the CSE contributions
with increasing wavelength. A more complete coverage
of the observed (u,v) plane is required to disentangle
this degeneracy.

%

%

\section{Validity of the proposed model}

Since the five collected baselines cover almost the whole (u,v)-plane,
we can safely claim that
we have detected the brightest clump in the dusty CSE of RY~Sgr.
However, the geometry of this CSE could be more complex
than described above. We therefore discuss here the effects 
 on the visibility curve of some departures in the proposed model.

First, the smoothness of the visibility curves leads us to discard 
the hypothesis of a dusty environment filled by several more
or less bright clumps. 
Indeed, any other heterogeneities in the CSE
would contribute with rather small perturbations to visibilities,
slightly changing the shapes and levels of the curves.
In order to test this hypothesis, we thus analysed the
effects of the presence of another unresolved clump onto our best model.
We then estimated in which conditions it could be
confidently detected with the present dataset.
Assuming the values estimated in Sect.~\ref{secres} for the stellar flux and 
for the CSE flux and size, we found that none of the
following heterogeneities would have been distinguishable in addition to the
already detected cloud:
(i) any cloud closer than typically 3-4\,mas from the star because
of the limited projected baselines ($\sim$130~m), 
(ii) any cloud closer than typically 3-4\,mas around the main clump at 
the same PA, or (iii) any
 clump fainter than $\sim$1-2\% of the 
total flux and located at a typical distance of 5-60~mas from the central star
(depending on PA). We also point out that the circumstellar layers located
beyond 60~mas are too extended in order to be efficiently explored with the MIDI
instrument.


As another verification,
we can also see in Fig.~\ref{figvismono} that the visibility curves $e$ and $f$ with the lower spatial frequency range
are almost straight and smooth and that the sinusoidal contribution is noticeable only at larger frequencies.
Thus, any bright cloud located at a larger separation than estimated before
would produce, in these two visibility curves,
a sinusoidal modulation that is not observed.
This would also not reproduce the relatively smoothed
shape of the visibility curves $a$ and $b$.
Therefore, we are confident in claiming that 
the contribution of an unique cloud
as described in Sect.~3 well fits simultaneously the five visibility
curves.
Any contribution from any other structure must be much fainter
than the main cloud already detected.

The strongest departure of the model with the 
observed visibilities is seen in curve $g$, the only dataset recorded in May 2005.
We investigated whether this departure could be
explained by a displacement of the cloud.
Considering a typical escape velocity of
275~km~s$^{-1}$ (Clayton et al.~\cite{cla03}) for RY~Sgr
and a distance of about 1.9~kpc (see Paper~1),
we estimate that this cloud 
could have moved radially by about 2-3~mas in one month but we found that such a displacement 
does not strongly affect the theoretical visibility curves.
Another option is to add to the geometrical model
a second cloud with about 5\% of the total flux,
at a separation of about 30~mas from the central star,
and a PA of about 135\degr. This putative cloud has to be
fainter than about 2\% one month later in order to be compatible
with the other visibility curves.
This may indicate a large dilution of the cloud but
the lack of data at high spatial frequencies
for PA~$>$~90\degr \ does not allow us to clearly verify that possibility.

Finally, we point out that the inclusion of any 
additional features to the geometrical model
could certainly help to better adjust the shapes 
of the theoretical visibility curves
but without providing more precise information. Indeed, 
the more we increase the complexity of a geometrical model
the more degenerated its parameters remain.
In any case, any other features that could be present in the CSE of RY~Sgr
would probably be faint or very close to the central star,
contributing only by small perturbations to the visibility curves.

\section{Discussion}
The collected VLTI/MIDI observations
are well interpreted with the simple geometrical model of the dusty 
environment of RY\,Sgr described in Sect.~3. 
Owing to this unprecendented study, we can claim that we have 
explored with a dynamic range better than 20 the inner 60~mas of the  
RY~Sgr environment. That corresponds to about 110~AU (following Paper~1,
we assume a distance of about 1.9~kpc and a photospheric angular radius 
of $\sim$0.15~mas for the central star).
 We can estimate that the CSE has a FWHM of about 120~R$_*$, or $\sim$35~AU and the detected cloud lies at about 100~R$_*$ from the central star ($\sim$ 30~AU).
This is the closest dusty cloud ever detected around a R~CrB-type variable
since the first direct detection with the ESO/NACO instrument
(see Paper~1).
However, such a distance is still too large to disentangle
between the two different scenarios  proposed for the formation
location of dusty clouds around R~CrB variables, either at
$\sim$2~R$_*$ or $\sim$20~R$_*$ from the central star. Interferometric observations with larger baselines 
and at smaller wavelength could
help to settle this issue. Using longer baselines in mid-IR is not an easy 
task: RY~Sgr can safely be observed with the VLTI 1.8m Auxiliary Telescopes 
with baselines up to $\sim$50m, then the correlated flux drops below the MIDI 
sensitivity limit. Observing at shorter wavelengths with the VLTI/AMBER 
near-IR recombiner appears to be a better solution. The clouds close to the 
star should be hotter, improving slightly the contrast, the spatial resolution 
is strongly increased and the accuracy better than in the N band. Moreover,
the closure phase provided by the use of three telescopes simultaneously is a 
powerful additional constraint, helping for the time monitoring of this 
kind of objects.

Moreover, assuming an improbable maximum value of about 275~km~s$^{-1}$
for the velocity projected on the sky of the detected cloud,
one can estimate that its ejection occurred more than $\sim$6~months before 
the epoch of the observations.
We found in the AAVSO\footnote{http://www.aavso.org/} light-curves of RY~Sgr that between early 2002 and the time of our observations two dimming events
occurred about 8~and 6~months before the VLTI observations.
Their durations were around 40~days and 4~months, 
and their recovering to maximum light took about 10~days and 2--3~months, 
respectively.
The cloud detected with MIDI was probably not one of those responsible for 
the dimming reported by AAVSO since this would require 
 too fast a displacement between the line-of-sight (epoch
of the minimum of brightness in the optical) and its
location at the date of the MIDI observations.
The detected cloud could, however, be related to a series of
ejections that produced the dimmings seen in the AAVSO light-curves.
With such an hypothesis, R~CrB-type variables  could
experience intense periods of material ejection. And, only part of the 
lost matter was up to now detected during a dimming episode. 

Finally, we emphasize that the observations presented here represent a
snapshot obtained within one month, June 2005. We still do not know how
the detected structures evolve with time, what is their radial velocity 
compared to the one of the dusty wind and how long are they steady. 
Since dust clouds are detected rather far from 
the central star, we suspected in Paper~1 that they are steady over periods 
of a few years.
They probably move away from the central
star leading to less obscuration of the stellar surface  and the return
to normal light would then not be caused by the evaporation of the clouds
close to the stellar photosphere as it has been suggested. 
Time series of visibility curves collected 
over several months could give crucial informations about
any displacement of the heterogeneities found around
R~CrB variables. This would definitively prove that (i) a dimming event 
would be related to an ejection of a dusty cloud on the line-of-sight
and to a sporadic ejection of stellar material towards any other direction
and, (ii) the duration of the return to maximum of brightness would
simply result from
the displacement of a cloud away from the line-of-sight.

\begin{acknowledgements}
We are grateful to the variable star observations from the AAVSO
International Database, contributed by observers worldwide and used
in this research.
\end{acknowledgements}

\end{document}